\documentclass[pra,letterpaper,twocolumn,showpacs,superscriptaddress,floatfix,nofootinbib]{revtex4}

\usepackage{graphicx,psfrag,amsmath,amssymb,amsfonts,bbm,latexsym,color,dcolumn,bm}

\begin{document}

\title{Proton radius puzzle and quantum gravity at the Fermi scale}

\author{Roberto Onofrio}

\email{onofrior@gmail.com}
 
\affiliation{Dipartimento di Fisica e Astronomia 'Galileo Galilei', 
Universit\`a di Padova, Via Marzolo 8, Padova 35131, Italy}

\affiliation{ITAMP, Harvard-Smithsonian Center for Astrophysics, 
60 Garden Street, Cambridge, MA 02138, USA}
\date{\today}

\begin{abstract}
We show how the ``proton radius puzzle'' emerging from the 
measurement of the Lamb shift in muonic hydrogen may be solved by
means of a binding energy contribution due to 
an effective Yukawian gravitational potential related to charged weak interactions.
The residual discrepancy from the experimental result should be mainly attributable 
to the need for the experimental determination of the gravitational radius of the proton.
The absence of an analogous contribution in the Lamb shift of electronic hydrogen should 
imply the existence of generation-dependent interactions, corroborating previous proposals.
Muonic hydrogen plays a crucial role to test possible scenarios for 
a gravitoweak unification, with weak interactions seen as 
manifestations of quantum gravity effects at the Fermi scale.
\end{abstract}


\maketitle
Hydrogen spectroscopy has played a major role for understanding the microscopic 
world in terms of quantum mechanics and quantum field theory \cite{Reviewspectroscopy}. 
Detailed studies of hydrogen have now reached an accuracy level limited by the proton 
size, expressed as the root-mean square charge radius $r_p=\langle r_p^2 \rangle^{1/2}$. 
To test quantum electrodynamics at the highest precision level, the proton 
size should then be determined with high precision from independent experiments. 
In the analysis of electron scattering experiments, a value of $r_p=(0.897 \pm 0.018)$ 
fm has been determined \cite{Sick,Blunden}. 
Higher precision determinations are possible using muonic hydrogen \cite{Boriereview}. 
Since the more massive muon has a smaller Bohr radius and a more significant 
overlap with the proton, the correction due to the finite size of the latter 
is more significant than in usual hydrogen. 
However, a recent measurement \cite{Pohl} reported a value of $r_p=(0.84184\pm 0.00067)$ 
fm, which differs by seven standard deviations from the CODATA 2010 value of 
$(0.8775\pm 0.0051)$ fm, obtained by a combination of hydrogen spectroscopy 
and electron-proton scattering experiments. 
Barring back-reaction measurement effects on the proton radius due to the use 
of a different leptonic probe, the radius of the proton is expected to be an 
invariable, constant quantity, even considering the underlying assumed lepton 
universality for electromagnetic interactions. 
This has  generated what is called the ``proton radius puzzle'' \cite{Pohlreview,Antogninireview}.
If the proton radius is kept at its CODATA value, this anomaly can be rephrased as if 
there is an excess of binding energy for the 2s state with respect to the 2p state equal to 
$\Delta E_{2s2p}=0.31$ meV, a $0.15 \%$ discrepancy on the Lamb shift theoretical expectation. 

This anomaly has elicited a number of theoretical hypotheses, including some invoking 
new degrees of freedom beyond the standard model \cite{Barger1,Tucker,Batell,Barger2}. 
In this context, pioneering papers have already discussed high precision 
spectroscopy as a test of extra-dimensional physics \cite{Arkani,Randall} both for hydrogen 
\cite{Luo1,Luo2}, helium-like ions \cite{Liu}, and muonium \cite{ZhiGang}, 
giving bounds on the number of extra-dimensions and their couplings. 
Values for the coupling constant to explain the proton radius puzzle in 
extra-dimensional models necessary were determined in a recent paper \cite{Wang}. 
The idea that extra-dimensions may be in principle tested with atomic physics tools 
is appealing also considering the paucity of viable experimental scenarios to 
test quantum gravity \cite{Ellis,Amelino1,Amelino2}. 
However, it would be most compelling to have a setting in which this
may be achieved in an economic fashion, that is without necessarily
introducing new free parameters conveniently chosen to accommodate 
{\it a posteriori} the experimental facts. 

In this letter, we try to provide such an approach through a tentative 
unification between gravitation and weak interactions conjectured in \cite{Onofrio}. 
We first discuss an effective potential energy between two pointlike 
masses which recovers Newtonian gravity at large distances while morphing
into an inverse square-law interaction with strength equal to the one of 
weak charged interactions at the Fermi scale, coinciding with the Planck scale. 
We then generalize this gravitational potential to the case of an extended 
structure like the proton, and evaluate the gravitational contribution to 
the Lamb shift using perturbation theory. 
The predicted contribution to the Lamb shift in muonic hydrogen is below 
the observed value within a factor of three, {\it i.e.} we derive 
$\Delta E_{2s2p}=0.106$ meV versus the experimentally
determined value of $\Delta E_{2s2p}=0.31$ meV. 
One potential source of discrepancy between our prediction and the
experimental result is then discussed in more detail.
This is then followed by a qualitative discussion of the possible
nature of the Yukawian potential of gravitoweak origin, including 
its selectivity towards the flavor of the fundamental fermions.
Finally we stress that a more accurate evaluation calls for the measurement 
of the gravitational radius of the proton, which is expected to significantly 
differ from the charge radius due to the gluonic energy density distribution 
for which no experimental access seems available. 

While we refer to \cite{Onofrio} for more details, we briefly recall
here that the idea we have explored is that what we call weak interactions, at 
least in their charged sector, should be considered as empirical manifestations 
of the quantized structure of gravity at or below the Fermi scale. 
This opens up a potential merging between weak interactions and gravity at the
microscale, a possibility supported by earlier formal considerations
on the physical consequences of the Einstein-Cartan theory \cite{Hehl}. 
Various attempts have been made in the past to introduce gravitoweak unifications 
schemes \cite{Hehl1,Batakis1,Batakis2,Loskutov,Alexander}, and 
a possible running of the Newtonian gravitational constant in purely four-dimensional 
models has been recently discussed \cite{Capozziello,Calmet}. 
The conjecture discussed in \cite{Onofrio} relies upon identification of a quantitative 
relationship between the Fermi constant of weak interactions $G_F$ and a renormalized 
Newtonian universal gravitational constant $\tilde{G}_N$ 
\begin{equation}
G_F=\sqrt{2}\left(\frac{\hbar}{c}\right)^2 \tilde{G}_N.
\label{Eq1}
\end{equation}
This expression holds provided that we choose $\tilde{G}_N=1.229
\times 10^{33} G_N=8.205 \times 10^{22}~\mathrm{m^3~ kg^{-1}~s^{-2}}$. 
Equation (\ref{Eq1}) differs from Eq. 2 in \cite{Onofrio} since we have adopted 
in this letter a more rigorous definition of Planck mass as the one 
corresponding to the equality between the Compton wavelength and the 
Schwarzschild radius, {\it i.e.} $\hbar/(M_P c)=2 G_N M_P/c^2$, the 
factor of 2 in the Schwarzschild radius having been omitted in the 
first analysis presented in \cite{Onofrio}. 

The identification of the Fermi constant with a renormalized Newtonian universal 
constant via fundamental constants $\hbar$ and $c$ allows to identify Fermi and 
Planck scales as identical, $\tilde{E}_\mathrm{P}=v$, where $\tilde{E}_\mathrm{P}$ 
and $v$ are respectively the renormalized Planck energy and the vacuum expectation value 
of the Higgs field, sometimes called the Fermi scale, avoiding then any hierarchy issue.
As discussed in \cite{Onofrio}, there are also a number of possible tests of this conjecture
that can span a wide range of energy, from the ones involved in the search for gravitational-like 
forces below the millimetre range \cite{Onofrio1,Antoniadis}, to the ones explored at the 
Large Hadron Collider, with the spectroscopy of exotic atoms in between. 
We have envisaged in \cite{Onofrio} the possibility that muonic hydrogen 
provides a suitable candidate, and this is now discussed in detail evaluating 
the generalized gravitational contribution to the Lamb shift in muonic hydrogen. 

First, we need to interpolate between the two regimes of weak gravity at macroscopic 
distances and the conjectured strong gravity/weak interactions at the microscale. 
Newtonian gravitation at large distances (with coupling strength $G_N$) can morph 
into weak interactions corresponding to a renormalized universal gravitational constant 
$\tilde{G}_N$ at small distances by means of a generalized potential energy 
$V_\mathrm {eff}$ for the gravitational interaction between two pointlike 
particles of mass $m_1$ and $m_2$
\begin{equation}
V_\mathrm{eff}(r)=-\frac{G_N m_1 m_2}{r} \left[1+\left(\frac{\tilde{G}_N}{G_N}-1\right)
e^{-r/\tilde{\Lambda}_\mathrm{P}}\right],
\label{Eq2}
\end{equation}
where we have assumed as Yukawa range the renormalized Planck length
$\tilde{\Lambda}_\mathrm{P}=\sqrt{2\hbar \tilde{G}_N/c^3}=8.014 \times 10^{-19}$~ m. 

Equation (\ref{Eq2}) reduces to ordinary gravity for $r>>\tilde{\Lambda}_\mathrm{P}$, whereas in 
the opposite regime of $r<< \tilde{\Lambda}_\mathrm{P}$ continues to have a $1/r$ 
behaviour, but with coupling strength proportional to $\tilde{G}_N$. 
In both limits, the evaluation in perturbation theory of the average gravitational 
energy should give no difference between the 2s and 2p states since any $1/r$ potential 
is degenerate for states with the same principal quantum number and different angular momenta. 
Therefore the difference we may evidence in this analysis will reflect the genuine 
deviation from a inverse square law characteristic of Yukawian potentials. 
To simplify the notation, in the following equations we will consider the 
parameter $\alpha=\tilde{G}_N/G_N-1$ and $\lambda=\tilde{\Lambda}_\mathrm{P}$,
as customary in the analysis of Yukawian components of gravity \cite{Onofrio1,Antoniadis}.
Notice that the second term in the righthandside of
Equation 2 becomes comparable to the first one at a length scale $R$ such
that $\tilde{G}_N/G_N e^{-R/\tilde{\Lambda}_\mathrm{P}} \simeq 1$,
{\it i.e.} $R \simeq \tilde{\Lambda}_\mathrm{P} \ln(\tilde{G}_N/G_N)\simeq
76 \tilde{\Lambda}_\mathrm{P}$. Thus the effect of the Yukawian component 
may be evidenced at length scales much larger than the
renormalized Planck length  $\tilde{\Lambda}_\mathrm{P}$, depending on the
precision available in the specific experimental scheme used to test
quantum gravity effects of this nature.

The evaluation of the gravitational energy between two particles is quantitatively 
different if one of them has an extended structure, as in the case of the proton.
We schematize the proton as a spherical object with uniform mass density
only within its electromagnetic radius $R_p$ related to the rms charge
radius through $R_p=\sqrt{5/3} r_p$ assuming a uniform charge density
inside the proton, $\rho_p=3m_p/(4\pi R_p^3)$.
With this assumption, the evaluation of the Newtonian potential
energy - the first term in the righthand side  of Eq. (\ref{Eq2}) - between 
the proton and a generic lepton of mass $m_\ell$ ($\ell=e, \mu, \tau$) yields
 
\begin{eqnarray}
V_{N\ell}(r)&=&G_N \frac{m_\ell m_p r^2}{2R_p^3}-\frac{3}{2}G_N \frac{m_\ell
  m_p}{R_p} \hspace{0.5cm} (0<r<R_p), \nonumber \\
V_{N\ell}(r)&=&-G_N \frac{m_\ell m_p}{r} \hspace{2.8cm} (r>R_p). 
\end{eqnarray}

The calculation of the energy contribution due to the Newtonian potential may 
be performed by means of time-independent perturbation theory applied to 2s and 2p states. 
Due to space isotropy we focus only  on the radial (normalized) components of 
the unperturbed wavefunctions which are, respectively

\begin{eqnarray}
R_{2s}(r) &=&  \frac{1}{(2a_\ell)^{3/2}}\left(2-\frac{r}{a_\ell}\right)
e^{-\frac{r}{2a_\ell}}; \\ 
R_{2p}(r) &=& \frac{1}{\sqrt{3}(2a_\ell)^{3/2}}\frac{r}{a_\ell}e^{-\frac{r}{2a_\ell}},
\end{eqnarray}
where $a_\ell$ is the Bohr radius, which also takes into account the
reduced mass of the lepton-proton bound system. 

\begin{figure}[t]
\includegraphics[width=1.00\columnwidth,clip]{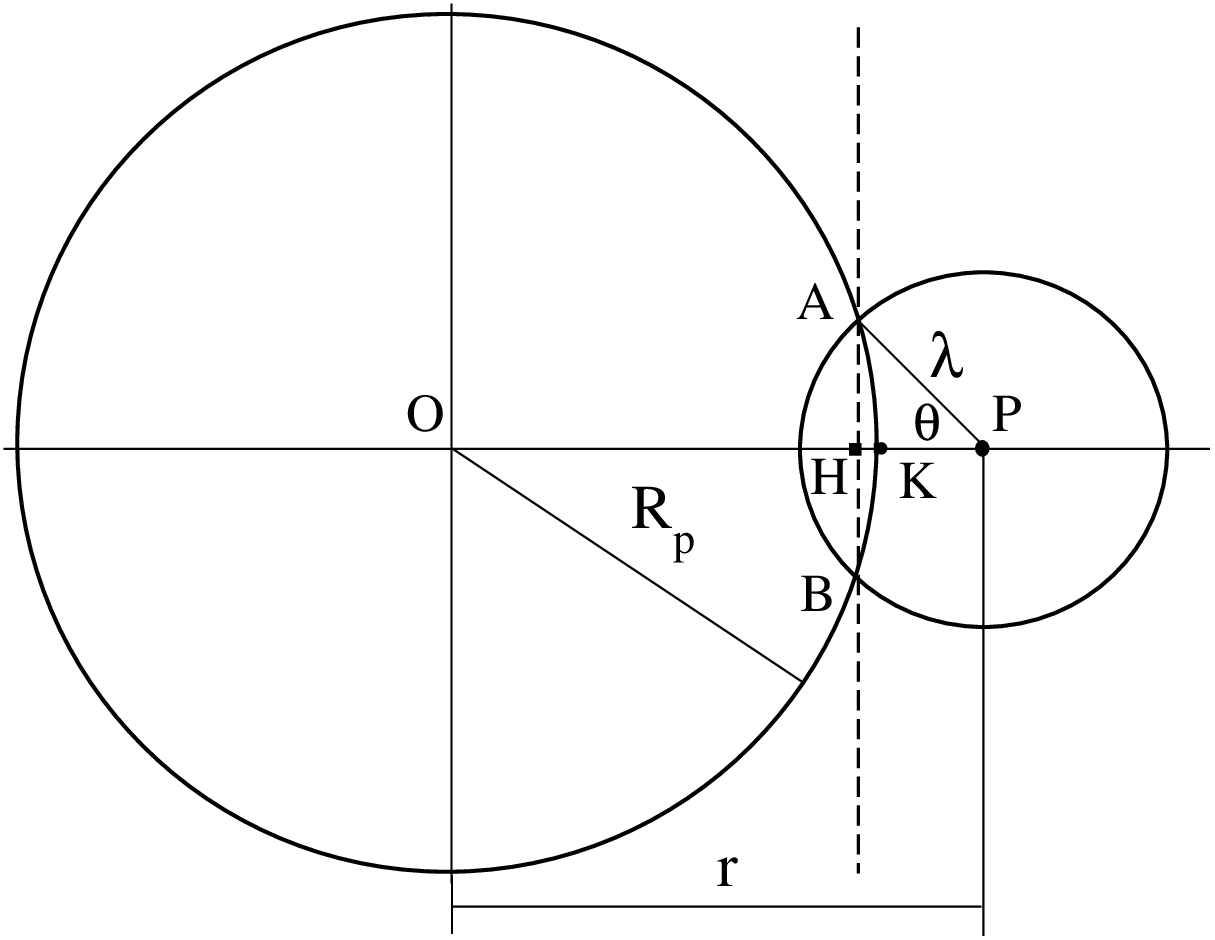}
\caption{Drawing (not to scale) of the relative proton and lepton
coordinates for the evaluation of the Yukawian potential energy term due to the 
extended structure of the proton. The circle on the left side centered
in point O represents the extent of the electric charge distribution
of the proton, assumed uniform inside the sphere of radius $R_p$. 
The circle on the right side represents the Yukawa range surrounding 
a lepton located at point P.}
\label{muonichlettfig1}
\end{figure}

By simple algebraic manipulations we obtain a compact expression for the 
Newtonian potential energy difference between the 2s and the 2p states as

\begin{eqnarray}
& & \Delta \langle V_{N\ell} \rangle_{2s2p}=\langle V_{N\ell} \rangle_{2s}-\langle
V_{N\ell} \rangle_{2p}=\frac{G_N m_\ell m_p}{R_p} \times \nonumber \\
& & \left[6(e^\beta-1)\beta^{-2}-6\beta^{-1}-3-\beta-\frac{\beta^2}{4}\right]e^{-\beta},
\end{eqnarray}
where we have introduced $\beta=R_p/a_\ell$.
It is easy to see that, in the $\beta\rightarrow 0$ limit of a
pointlike proton, 
$\Delta \langle V_{N\ell}\rangle_{2s2p}\rightarrow G_N m_\ell m_p \beta^3 e^{-\beta}/(20 R_p)
\rightarrow 0$. For a proton of finite size, this contribution is
different for hydrogen and muonic hydrogen since the gravitational
mass $m_\ell$ appears both as a factor and in the expression for the
Bohr radius $a_\ell$ on which $\beta$ depends. We notice also that
this contribution tends to increase the energy of the 2s state, as this 
state is more sensitive to the finite size of the proton with respect 
to the 2p state, analogously to the case of the Coulombian attraction.
Moreover, it is easy to check that the difference is absolutely
negligible in our context, being about 37 orders of magnitude smaller 
than the experimentally observed anomaly in muonic hydrogen, so it
cannot play any role in its understanding. This is also consistent 
with a simple estimate obtained just considering a point-like proton. 
From the estimate of the absolute Newtonian potential energy 
$V_N \simeq G_N m_{\ell} m_p/a_{\ell}$ we obtain a value 
of $V_N \simeq 4.6 \times 10^{-34}$ eV for muonic hydrogen. 
The presence of an extended proton structure and the fact that 
the Lamb shift contribution is given by the difference between 
the potential energy in the 2s and 2p states further suppress the 
expected Newtonian contribution. However, this shows that by 
boosting the gravitational term using $\tilde{G}_N$ instead of
$G_N$, {\it i.e.} by more than 33 orders of magnitude, the value 
of the estimate leads to $V_N \simeq 565$ meV. 

Based on this promising estimate, we now evaluate the Yukawian 
component, {\it i.e.} the second term in the righthand side of Eq. (\ref{Eq2}). 
The first step is the evaluation of the effective potential due to 
the interaction of the muon with an extended proton. 
Assuming a semiclassical picture, we expect that if the muon is outside 
the proton by at least an amount $\simeq \lambda$ there will be a negligible 
Yukawian gravitational contribution, and the same will occur if the muon is 
completely inside the proton, since the uniform distribution of the proton 
mass will exert isotropic interactions averaging out to zero. 
A non-zero value for the proton-lepton interaction instead occurs while the muon is partially 
penetrating inside the proton radius within a layer of order $\lambda$, reaching its 
maximum intensity when the muon is located at the proton radius, {\it i.e.} when 
point P coincides with point K in figure 1. 
The exact calculation for the Yukawian potential felt by the lepton should proceed 
by integrating the Yukawian potential contributions due to each infinitesimal volume 
element inside the proton. 
This calculation involves exponential integral functions and is not trivially 
performed analytically. 
A brute force numerical integration is  also subtle since the problem involves length scales, 
like $\lambda$, $R_p$, and $a_{\ell}$, differing by several orders of magnitude. 
We then proceed with an analytical evaluation by introducing two approximations. 
First, we truncate the Yukawian potential between two pointlike masses 
at distance $r$ in such a way that
$V_{Y\ell}(r) = - \alpha {G}_N m_\ell m_p/r$ if $0<r<\lambda$
and $V_{Y\ell}(r)=0$ if $r>\lambda$.
This means that the evaluation of the potential is carried out only in the 
region of intersection between two spheres, one of radius equal to $R_p$, 
the other of radius equal to the Yukawa range $\lambda$. In using this 
approximation, the amplitude of Yukawian potential is then overestimated 
in the $0<r<\lambda$ region, while it is underestimated in the region 
corresponding to $r>\lambda$.
Second, we assume that the intersection region is a spherical cap, 
an approximation corresponding to $R_p >>\lambda$, 
which we have found {\it a posteriori} well satisfied for the 
value of $\lambda$ able to justify the anomalous Lamb shift.  

The potential felt by the lepton at a distance $r$ from the proton center 
is then evaluated by integrating the infinitesimal potential energy 
contributions over the volume of the spherical cap of the 
sphere of radius $\lambda$ centered on the lepton location. 
With reference to figure 1, and using spherical coordinates for the 
infinitesimal volume $dv$, this leads to a Yukawian potential energy as follows

\begin{eqnarray}
V_{Y\ell}(r)&=&-2\pi \alpha G_N m_\ell \rho_p \lambda^2, \nonumber \\
V_{Y\ell}(r)&=&\pi \alpha G_N m_\ell \rho_p [(r-R_p)^2+2\lambda(r-R_p)-\lambda^2], \nonumber \\
V_{Y\ell}(r)&=&- \pi \alpha G_N m_\ell \rho_p [(r-R_p)^2-2\lambda(r-R_p)+\lambda^2], \nonumber \\
V_{Y\ell}(r)&=&0,
\end{eqnarray}
respectively in the four regions $[0,R_p-\lambda]$, $[R_p-\lambda,R_p]$, $[R_p,R_p+\lambda]$, and
$[R_p+\lambda,+\infty[$, with continuity of the Yukawa potential enforced for all boundaries. 
As discussed above, this potential corresponds to a net attractive force only in the range 
$R_p-\lambda<r<R_p+\lambda$.

\begin{figure}[t]
\includegraphics[width=1.0\columnwidth,clip]{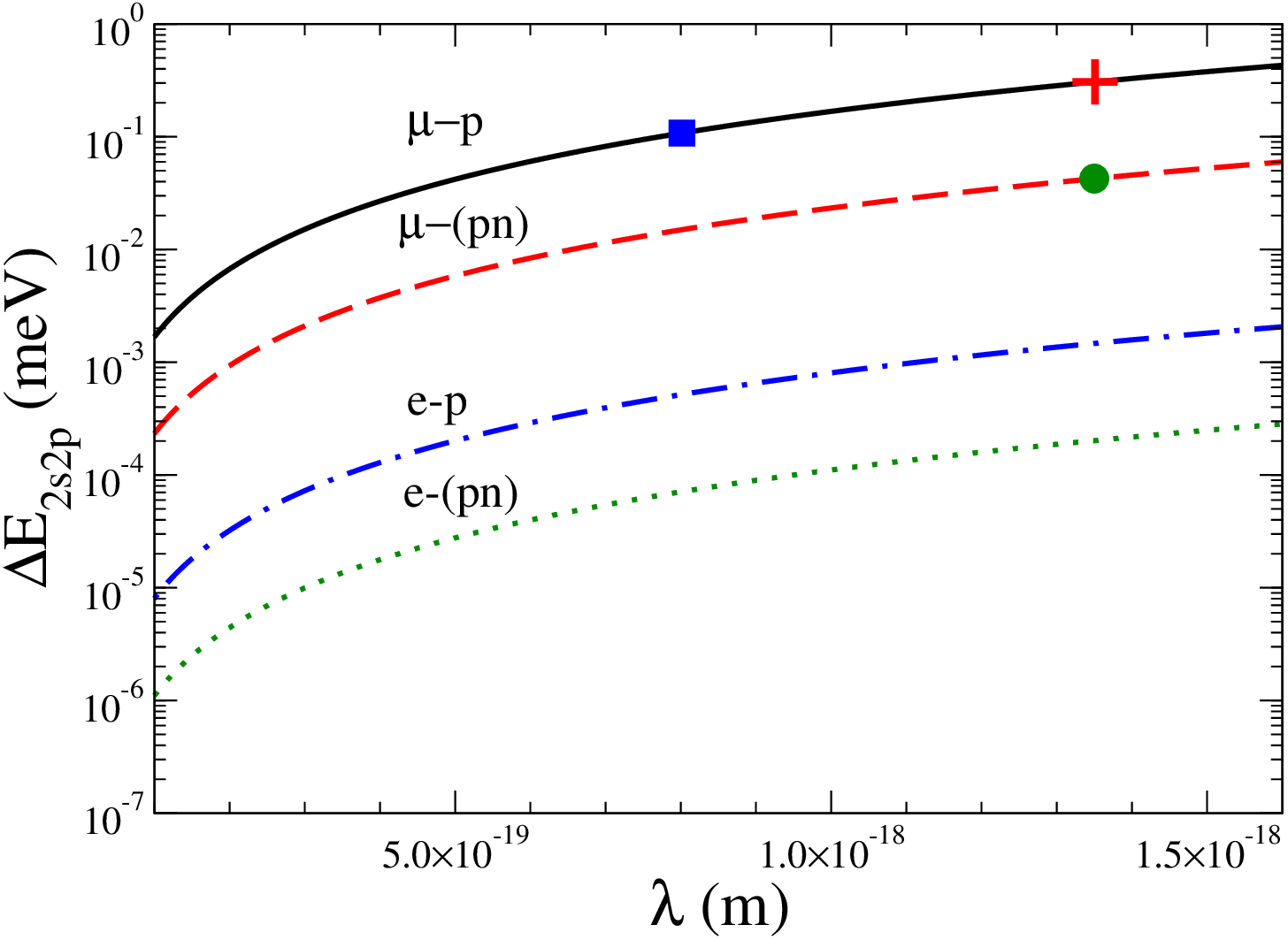}
\caption{Dependence of the calculated gravitational energy difference between the
2s and 2p energy levels $\Delta E_{2s2p}$ for various bound states,
versus the Yukawa range of short-distance gravity. 
From top to bottom, the curves are relative to the cases of muonic 
hydrogen (solid black line), muonic deuterium (dashed red line), hydrogen 
(dot-dashed blue line), and muonic deuterium (dot green line), assuming 
for the proton and deuteron radii their CODATA 2010 value.
The experimental value for the muonic hydrogen case $\Delta E_{2s2p}=$
0.31~ meV is shown (red cross), which is explained assuming a value of 
$\lambda=1.35 \times 10^{-18}$ m at the effective coupling strength 
and proton radius assumed above. The prediction of the model for 
$\lambda=\tilde{\Lambda}_\mathrm{P}$, corresponding to $\Delta E_{2s2p}$=0.106 meV 
(blue square), is also shown. The prediction of the anomaly for muonic
deuterium at the same $\lambda$ accommodating the anomaly for muonic
hydrogen is 42.6 $\mu$eV (greeb dot).}
\label{muoniclettfig2}
\end{figure}

\begin{figure}[t]
\includegraphics[width=1.0\columnwidth,clip]{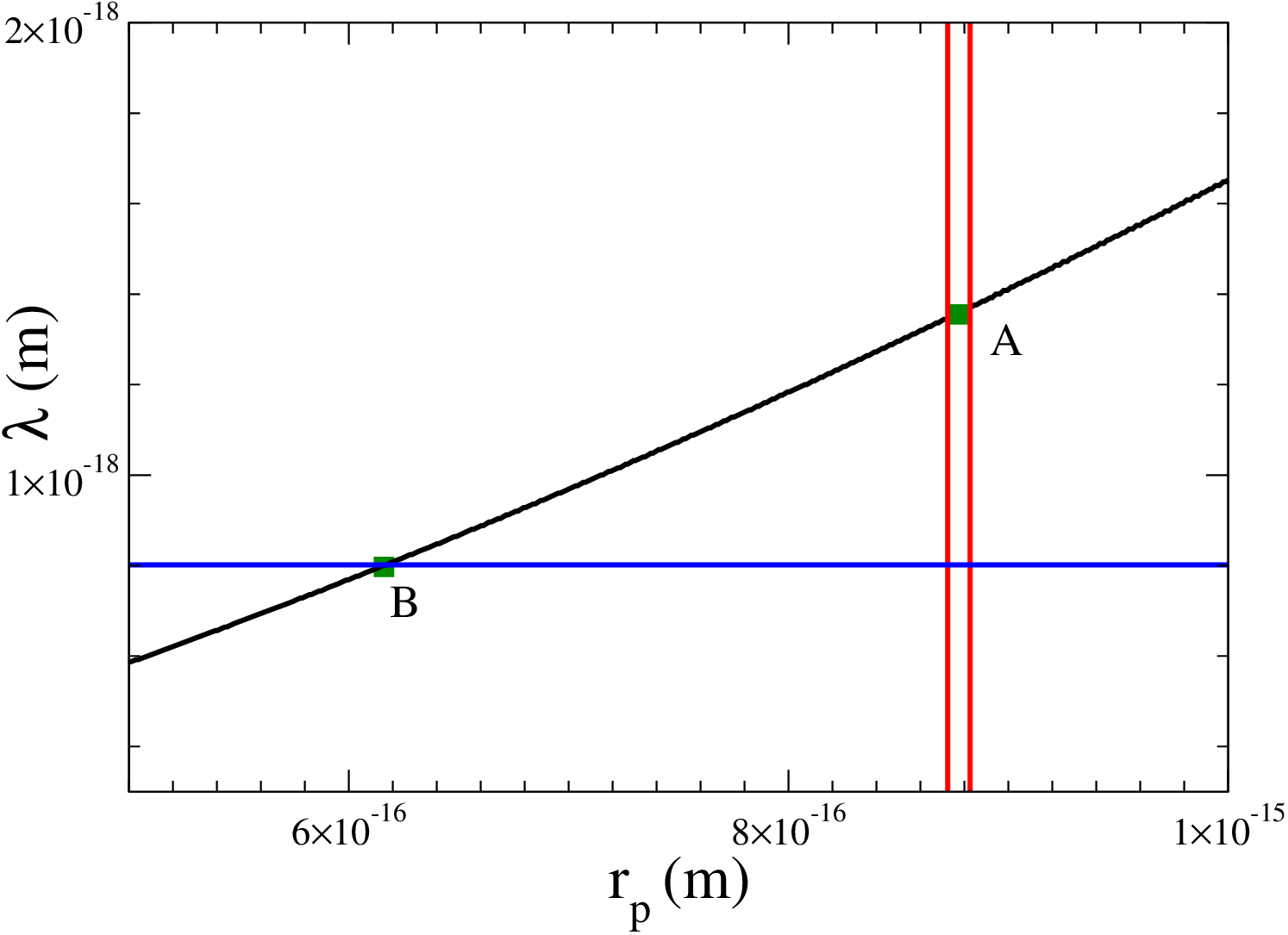}
\caption{Locus in the ($\lambda,r_p$)-plane of all solutions to proton radius 
puzzle, giving rise to a Lamb shift excess 0.309 meV $\leq\Delta E_{2s2p}\leq$ 
0.311 meV, as shown by the monotonically increasing curve. 
The range of values allowed, within one standard deviation, for the proton 
radius according to the CODATA value are delimited by the two vertical lines, 
the horizontal line corresponding instead to $\lambda=\tilde{\Lambda}_\mathrm{P}$. 
The experimental value for the anomalous Lamb shift is reproduced for instance by 
choosing $\lambda \simeq 1.7 \tilde{\Lambda}_\mathrm{P}$ while having 
$r_p$ at its CODATA value (solution A), or by choosing the gravitational radius
of the proton equal to $\simeq$ 0.7 times the CODATA value for the
charge radius while keeping $\lambda=\tilde{\Lambda}_\mathrm{P}$ (solution B).}
\label{muoniclettfig3}
\end{figure}

We evaluate the expectation value of the Yukawian potential energy in the 2s and 2p 
states, $\langle V_{Y\ell} \rangle_{2s}$ and $\langle V_{Y\ell} \rangle_{2p}$, according 
to perturbation theory. 
The Yukawian component of the gravitational potential is still much smaller that the 
Coulombian potential even if it is coupled through $\tilde{G}_\mathrm{N}$ at short 
distances, since $\tilde{G}_\mathrm{N} m_\ell m_p <<e^2/(4\pi\epsilon_0)$. 
The difference between the two contributions gives, after lenghty
algebraic simplifications, the expression
 
\begin{eqnarray}
& & \Delta \langle V_{Y\ell} \rangle_{2s2p}=
\langle V_Y \rangle_{2s}-\langle V_Y \rangle_{2p}=
-\frac{\alpha G_N m_\ell m_p}{16R_p^3} \times \nonumber \\
& & \left\{\lambda^2 f(y)+a_\ell^2\left[g_{-}(y,\beta)+g_{+}(\beta,z)\right]\right\}, 
\label{Eq8}
\end{eqnarray}
where $f(y)=2(2-y)y^3 e^{-y}$, $g_{\pm}(a,b)=\sum_{n=0}^6 h_{n}^{\pm} G_n(a,b)$,
$G_n(a,b)=a^n e^{-a}-b^n e^{-b}$, $y=(R_p-\lambda)/a_\ell$, $z=(R_p+\lambda)/a_\ell$, $v=\lambda/a_\ell$, and 
\begin{eqnarray}
& &  h_0^{\pm}= h_1^{\pm}=\pm 24(6 \mp v -\beta),
\nonumber \\
& &  h_2^{\pm}= 12(\pm 1-v \mp \beta),
\nonumber \\
& &  h_3^{\pm}=\mp 2(-12 \pm 2v+2\beta \pm v^2 \pm 2\beta v+\beta^2), \\
& &  h_4^{\pm}= 2\left(\pm 3+v \pm \beta+\frac{v^2}{2}+\beta v \pm \frac{\beta^2}{2}\right),
\nonumber \\
& &  h_5^{\pm}= -2(v\pm \beta), ~~h_6^{\pm}= \pm 1. \nonumber
\end{eqnarray}
The overall contribution is negative, {\it i.e.} $\Delta E_{2s2p}<0$, indicating
that the 2s state gets more bounded than the 2p state. 
This outcome differs from the case of the long-range Newtonian component since 
in the latter case the 2s state, exploring more the proton interior, gets a weaker 
binding, as only the inner mass is relevant for the gravitational potential. 
This feature is not shared by the Yukawian potential as it does not fulfil the 
Gauss theorem, only the proton mass nearby the lepton matters regardless of its 
location with respect to the proton center of mass. We will indicate from now on 
the absolute value of $\Delta E_{2s2p}$ with the implicit understanding that it is negative.

The result of this analysis is shown in figure 2, where $\Delta E_{2s2p}$ is plotted 
as a function of the Yukawa range $\lambda$ for a value of $\alpha$ corresponding to $\tilde{G}_N$. 
For a value of $\lambda=\tilde{\Lambda}_\mathrm{P}$ we obtain a value of 
$\Delta E_{2s2p} \simeq$ 0.106 meV, about 2.8 times smaller than the measured value. 
The experimental value of $\Delta E_{2s2p}$ is instead obtained from Eq. (\ref{Eq8}) 
by assuming a value of $\lambda=1.35 \times 10^{-18} {\mathrm m} \simeq 1.7 \tilde{\Lambda}_\mathrm{P}$. 
We notice that the prediction $\Delta E_{2s2p}$ assuming first-principle parameters 
$\alpha=\tilde{G}_{\mathrm N}-1$ and $\lambda=\tilde{\Lambda}_\mathrm{P}$ as inspired 
by the conjecture in \cite{Onofrio} is remarkably close to the experimental value, in 
spite of the drastic assumption of uniform mass density for the proton inside its electric 
radius and of the two approximations used in the calculation.
This analysis is complemented by showing, in figure 3, the combined
effect of $\lambda$ and $r_p$ on the evaluation of $\Delta E_{2s2p}$. 
We have evaluated the points in the $(r_p,\lambda)$-plane which allow to obtain a value of 
$\Delta E_{2s2p}$ in the $(0.311,0.309)$ meV interval, see monotonically increasing line. 
The vertical band is the CODATA 2010 range of values for $r_p$, and the horizontal 
line corresponds to the value of $\lambda=\tilde{\Lambda}_\mathrm{P}$. 
Exact validation of our model corresponds to a single intersection point among the three 
curves, which is not achieved within 80 $\%$ in $\lambda$ and 40 $\%$ in $r_p$.

Like in the Newtonian case, it is possible to give a 'back-of-the envelope' estimate of 
the effect which is consistent with the lengthy calculation resulting in equation (\ref{Eq8}). 
By thinking of a muon trajectory as in a semiclassical, Bohr-Sommerfeld approach, we 
argue that the 2p state does not acquire basically any Yukawian contribution as the 
corresponding trajectory is far away from the proton radius. 
The muon in the 2s state instead penetrates inside the proton, and then the Lamb shift 
basically coincides with the shift expected by the 2s state alone. 
Due to the short-range nature of the Yuwawa potential, the estimate of the excess 
energy in the 2s state due to the Yukawian component will involve the Yukawa range 
and the effective gravitational mass of the proton effectively seen by the muon, in 
such a way that the absolute Yukawian potential energy in the 2s state is 
$V_Y \simeq \tilde{G}_N m_{\mu} m_{p}^{\mathrm{eff}}/\lambda$. 
The effective mass of the proton participating to the Yukawa interaction is 
only the one intercepted by the muon within its Yukawa range (in the truncation 
approximation we have adopted), $m_{p}^{\mathrm{eff}} \simeq (\lambda/R_p)^3 m_p$ so we get 
\begin{equation}
V_Y \simeq \tilde{G}_N \frac{m_{\mu}m_{p}\lambda^2}{R_p^3} \simeq 
~0.18~ {\mathrm{meV}},
\end{equation}
which is within a factor less than two from both the
comprehensive analytical evaluation and the experimental value of the anomaly.

All possible refinements of the calculations are limited by the fact that in 
this approach the knowledge of the mass density distribution is essential.
In particular, no information is currently available on the density distribution for the 
gluonic fields, the most important component of the proton at the level of determining its
mass. This, at the moment, seems the most critical issue preventing a more 
quantitative comparison of the model with the experimental result. 
It seems plausible that gluons, only sensitive to the attractive color interaction, tend 
to cluster more than valence quarks which are also sensitive to the electromagnetic interaction 
acting both attractively and repulsively depending on the quark flavors, thereby decreasing 
the effective gravitational radius of the proton below its electromagnetic value. 
A smaller gravitational radius could bring the prediction more in line with the 
experimental value assuming $\lambda=\tilde{\Lambda}_P$ as visible in figure 3.

In spite of the limitation arising from the lack of knowledge of the gravitational 
proton radius, it is worth proceeding with the analysis of other systems in which 
this putative Yukawian potential may also give rise to observable effects and predictions. 
As visible in figure 2, the expected contribution in muonic deuterium is suppressed 
since the approximate doubling of the gravitational mass of the nucleus cannot compensate 
for the cubic dependence of the larger deuteron radius.
By repeating the evaluation for the Lamb shift in (electronic) hydrogen, we obtain a value 
which is too large, since it corresponds to 0.52 $\mu$eV at $\lambda=\tilde{\Lambda}_\mathrm{P}$ and 
1.47 $\mu$eV at $\lambda=1.35 \times 10^{-18}$ m, if using the same CODATA 2010
proton radius. Basically, due to the fact that $a_\ell >> R_p, \lambda$ for both 
electrons and muons, the leading difference between electrons and muons is the direct effect 
of their gravitational mass, with the forms factors due to the extended structure of the 
proton expressed through the functions $f(y)$, $g_{\pm}(a,b)$ in Eq. (\ref{Eq8}) being quite similar. 
The gravitational contribution in electronic hydrogen is therefore 
suppressed with respect to the one of muonic hydrogen by their mass ratio.  
This is in principle a big issue for the validation of the proposed model. 
The Lamb shift in hydrogen is known with an extraordinary precision which 
cannot incorporate such a large contribution, corresponding to an anomalous 
frequency shift of about 0.2 GHz against an absolute accuracy of
9.0 KHz, or $8.5 \times 10^{-6}$ relative accuracy on the experiment-theory comparison. 

Among possible ways to go around this issue while continuing to pursue this approach is 
to assume that the effective interaction corresponding to the Yukawian part in equation 
(\ref{Eq2}) does not act among fundamental fermions belonging to the same generation. 
Such a flavor-dependent interaction naturally spoils the universality characteristic of gravitation. 
However, it should be considered that in a possible gravitoweak unification 
scheme, the emerging structure should presumably incorporate features of both 
weak and gravitational interactions.
The former interaction is manifestly flavor dependent, as shown in the 
presence of CKM and PMNS mixing matrices for the charged current, so 
it is not a priori impossible that the interaction corresponding to 
the Yukawa component in equation (\ref{Eq2}) is highly selective in flavor content.
Our assumption is aligned with recent attempts to justify the muonic hydrogen 
anomaly in terms of interactions differentiating between leptons, thereby 
violating their assumed universality \cite{Barger1,Tucker,Batell,Barger2}.
This solution could inspire searches for models in which a ``hidden sector'' 
of the standard model includes intermediate bosons of mass in the range of 
the Higgs vacuum expectation value mediating interactions which have mixed 
features between the usual charged weak interactions and gravitation, for 
instance heavier relatives of the $Z^0$ boson. A flavor-dependent interaction 
could also potentially contribute to the understanding of the mass difference 
between charged leptons, an unsolved puzzle since the famous question by Isidor Rabi. 

In conclusion, we have shown that a solution to the ``proton radius puzzle'' is
potentially available by means of an effective Yukawian potential originated 
by the morphing of Newtonian gravitation into weak interactions at the Fermi 
scale as conjectured in \cite{Onofrio}, without additional parameters with 
respect to the ones already present in the standard model.  
The electric charge distribution of the proton investigated by using leptonic 
probes is not accurately representative of the mass distribution, due to 
the leading gluonic contribution to the proton mass. 
Furthermore, we have been forced to assume that the electron does not 
interact with the proton via the same effective Yukawian interaction 
as the muon, since the expected anomalous contribution in the former case 
is exceedingly large with respect to what is observed in the hydrogen Lamb shift. 
This partially spoils the requirement for simplicity and universality 
but it is also plausible in the light of the complex structure of weak 
interactions which are flavor and generation dependent. 
We believe that this approach could help guiding future experiments aimed 
at testing the lepton-hadron universality as in the proposed muon-proton 
scattering \cite{Gilman}, precision observables in muonium \cite{Oram,Badertscher,Woodle}, 
and Lamb shift measurements in deuterium \cite{Carboni,Martynenko,Krutov}.

\end{document}